\documentstyle{aipproc}

\renewcommand{\baselinestretch}{2}

\newcommand{\SS}{\renewcommand{\baselinestretch}{1}   \tiny \normalsize}

\begin{document}
\SS

\title{Bayesian Blocks: Divide and Conquer, MCMC, and Cell Coalescence Approaches}
\author{Jeffrey D. Scargle}
\address{NASA-Ames Research Center\\
       MS 245-3, Moffett Field, CA\footnote{Email: jeffrey@sunshine.arc.nasa.gov}}

\maketitle

\vskip 0.5 in

\noindent
{\large From the 19th International Workshop on 
Bayesian Inference and Maximum Entropy Methods 
(MaxEnt '99), August 2-6, 1999, 
Boise State University,
Boise, Idaho, USA.  
This conference proceedings will be
published (Josh Rychert, Editor).}

\vskip 0.5 in

\noindent
{\bf Key Words:} Poisson process, 
changepoint detection, 
time series, 
image processing, 
Voronoi tessellation

\begin{abstract}
Identification of local structure
in intensive data -- 
such as time series, 
images, and higher dimensional processes --
is an important problem in astronomy.
Since the data are typically generated by
an inhomogeneous Poisson process,
an appropriate model is one that 
partitions the data space
into cells, each of which is described by
a homogeneous (constant event rate) Poisson process.
It is key that the sizes and locations 
of the cells are determined by the data,
and are not predefined or even constrained to
be evenly spaced.
For one-dimensional time series,
the method amounts to 
Bayesian changepoint detection.
Three approaches to solving the
multiple changepoint problem
are sketched, based on:
(1) divide and conquer with single changepoints,
(2) maximum posterior for the number of changepoints,
and
(3) cell coalescence.
The last method starts from the 
Voronoi tessellation of the data,
and thus should easily generalize
to spaces of higher dimension.
\end{abstract}

\newpage
%\tableofcontents

\section{Introduction: The Data Analysis Problem}

Developments in detector technology for high energy 
astrophysics\footnote{The term {\it high energy astrophysics} 
is used loosely for both 
the study of astronomical objects which produce and emit
large amounts of energy, and for observations of radiation
consisting of high-energy photons, {\it e.g.} X-rays and
gamma-rays.  Often the two meanings coincide.}
have led to observation systems capable of reporting 
accurate arrival times for individual photons.
These times, while not binned, are quantized in 
microsecond-scale units I like to call ``ticks''
-- since they are in fact generated by the
ticking of the computer clock on-board the spacecraft.
In the approximation where the ticks are
short compared to time scales of interest,
we very accurately have a Poisson\index{Poisson} process. 
Note that, depending on the nature of the
variability of the process, different
mathematical models apply, as
indicated in the table below.

{\fboxsep=.01in \framebox{\parbox[t]{7in}{
\vspace{.2 in}
\hrule{} 
{\large
\vspace{-.02in}
\begin{tabbing}
%% set tab stops
\hspace*{.25in} \= \hspace*{2in} \= \hspace*{1in} \kill
\> Nature of       \> Mathematical \\
\> Variability         \> Model Process      \\
\end{tabbing}
\vskip -.2in
\hrule
\vskip -.4in
\begin{tabbing}
%% set tab stops
\hspace*{.25in} \= \hspace*{2in} \= \hspace*{1in} \kill
\> Constant             \> Homogeneous Poisson\index{Poisson}  \\
\>                \>    \\
\> Deterministic        \> Inhomogeneous Poisson \\
% \> ({\it e.g.} periodic)   \>    \\
\>                \>    \\
\> Random            \> Doubly Stochastic Poisson \\
\>                      \> (Cox Process) \\
\end{tabbing}
\vspace{-.2in} 
\hrule{} 
\vspace{.2 in}
}}}

\vspace{.2 in}

A number of mathematical 
references \cite{kutoyants,snyder,andersen,barndorff,stoyan}
describe the nature of spatial  Poisson\index{Poisson} 
processes.
Time series are the $1D$ special case, consisting of 
streams of numbers representing photon detection times.
In this context, the Poisson\index{Poisson} distribution 
is so accurate that it is hardly an approximation.
Statisticians seem to have a hard time understanding this point.  
I believe the culprit is the fact that the Poisson\index{Poisson} 
process is usually taught as the limit of a binomial process or the like.  
Technically, our data do comprise a
{\it finite Bernoulli lattice}\index{Bernoulli,lattice}\cite{stoyan},
since typically more than one event cannot be recorded at a given tick. 
(See \cite{christ}
for an interesting discussion of a lattice
theory of quantum fields.)
But the rates in units of events\footnote{We use the term {\it event}
for photon detections, or other data points.}
per tick are so low
that the probability of such multiple events is very low.

In practice, the most significant imperfection 
is a departure, not from the assumed distribution,
but from the assumption of independence of the process
at different times.
Photon detectors always have a finite {\it dead time} --
an interval after the detection of one event
during which another photon cannot be recorded,
either because the detector mechanism itself
is temporarily paralyzed, or the data system is
too busy processing the event.
This yields interdependences in the 
detection of photons close together in time.

The discrete nature of photon counting data is most often
considered a challenge.  Many analysts first
bin their data and then apply standard methods.
Further, the notion is rife 
that not only are bins necessary, 
but that they must be large and
contain enough events to 
produce a ``statistically significant sample.''
This is wrong and wasteful.
Analysis can
be carried out directly on the event data.
It is my belief that the discreteness 
and utter simplicity of the fundamental
event -- a photon was detected or it wasn't -- are a
big advantage.
This simplicity and the fact that 
the observational errors accurately 
follow a distribution with essentially no free 
parameters\footnote{The Poisson rate parameter is 
pretty closely nailed at the local event rate.\index{Poisson}}
mean that 
the posterior marginalization can be carried out exactly, 
at least for some of the nuisance parameters 
[see Eqs. (\ref{marg_like_1}) and \ref{marg_like_2}].

Happily, the simplicity of the basic events 
allows almost immediate generalization of 
one-dimensional results to data spaces of higher dimension.
Simple examples are cases where the photons, 
in addition to being timed, 
are also tagged with spectral and spatial information -- {\it i.e.} 
energies of photons and their location on the sky.
Processes that sample the density 
of events in spaces of various dimension (denoted $D$) include:
\begin{itemize}
\item[$\bullet$] time series\index{time series} ($D=1$)
\item[$\bullet$] other sequential data ($D=1$; {\it e.g.}, genetic sequences)
\item[$\bullet$] images ($D=2$)
\item[$\bullet$] time sequences of images ($D=3$)
\item[$\bullet$] time sequences of spectra ($D=2$)
\item[$\bullet$] points in parameter spaces (any $D$)
\end{itemize}

\noindent
The corresponding data types 
have in common that they comprise sets of points in some well defined space. 
Since the underlying process 
defines the intensity of some physical quantity over the space, 
we call these {\it intensive data}\index{intensive data}.  
The actual data can consist of a list of points,
or the number of points in pre-selected bins
({\it e.g.} image {\it pixels}), or other data modes.
A problem of broad interest is the detection of 
local\footnote{I use the term {\it local} to
distinguish features of limited extent
from {\it global} features, such as
periodic signals extending over the whole of the data space.}
structures in such data.  
Pulses, or other time-domain structures in 
time series\index{time series} data, 
features in images
and clusters in parameter spaces of 
various dimension (classification)
are examples.

For one-dimensional time series\index{time series},
the method of {\it Bayesian Blocks}\index{Bayesian blocks}
approximates the signal as 
a piecewise constant Poisson\index{Poisson} process \cite{scargle_v}.
Generalizing this representation to 
a space, denoted $S$, of arbitrary dimension,
the basic problem can be phrased as the following
relatively straightforward Bayesian 
{\it Maximum a posteriori}\index{Maximum a posteriori\cite{ruan}
 (MAP)} (MAP) problem:\\
{\fboxsep=.01in \framebox{\parbox[t]{5in}{
\begin{center}
Consider all partitions of $S$ into subvolumes, or {\it cells}.\\
Model the event rate within each cell as a homogeneous Poisson\index{Poisson} Process.\\
{\bf Among all such partitioned models, which is the most likely?} 
\end{center}}}
\vspace{.2 in}

\noindent

We determine the parameters in order of
how fundamental they are to the nature of the model.
The piecewise constant Poisson nature of the model
is assumed, so the most fundamental parameter is
$N_{cp}$, the number of changepoints.
This is determined by marginalizing all of the
other parameters -- the locations of the changepoints
and the Poisson rates between them.
Then the locations of the changepoints are
determined.
Finally, the block event rates are 
determined; trivially, the MAP value 
is just the ratio of $N$, the number of events in the cell,
to $V$, its volume.

The resulting evidence in favor of this model for a single cell,
called the {\it marginal likelihood}\index{marginal likelihood},
and defined below in Eq. (\ref{marg_like_1}),
is a simple function of $V$ and $N$ 
[below, Eqs. (\ref{marg_like_1}) and (\ref{marg_like_2}),
and also see \cite{scargle_v}].
For binned data and other data modes, and for other 
distributions than the Poisson\index{Poisson} one assumed here, the 
likelihood\index{likelihood} is similarly an explicit function of the 
same two parameters.  
We assume the cells are independent, 
so the total likelihood is the product of the likelihoods\index{likelihood} 
of the cells (see Section \ref{appr_3}).

\section{Exact Bayes Factors}

For the multiple changepoint\index{changepoint}
problem we need to evaluate the posterior probability
for piecewise constant models, given the data.
An important simplification is
that the marginal likelihoods\index{likelihood}
and posterior probabilities\index{posterior}
for such models factor into the product 
of the same quantities 
for each independent segment of the model.
We refer to these segments as {\it blocks}
in $1D$ contexts, and  as {\it cells} in higher dimensions.
This section gives the computation of the
posterior for a single cell, which
can then be used in various ways -- such as
the evaluation of Bayes factors\index{Bayes factors}
for comparison of two or more models.
The general form of 
Bayes factor\index{Bayes factors} comparing 
two models $M_{1}$ and $M_{2}$, given data $D$
\{{\it c.f.} \cite{gelman}, eq. (6.4)\} is:
\begin{equation}
\mbox{Bayes factor}(M_{2};M_{1}) =
{p(D|M_{2}) \over p(D|M_{1})}
= 
{\int p(\lambda_{2}|M_{2}) p(D|\lambda_{2},M_{2}) d\lambda_{2}
\over 
\int p(\lambda_{1}|M_{1}) p(D|\lambda_{1},M_{1}) d\lambda_{1}
}
\label{bayes_factor}
\end{equation}
\noindent
where $p(\lambda|M)$
is the prior on $\lambda$ and
$p(D|\lambda,M)$ is the likelihood.
Given the discussion above, 
it is easy to compute the appropriate factor
for a Poisson\index{Poisson}
model with rate parameter $\lambda$ (units: events per unit
volume) for a set of events in some block or cell.
The Poisson Likelihood, obtained by 
multiplying likelihoods for individual 
sample bins, is
\begin{equation}
L( N | \lambda, V ) =
e^{ - V\lambda } \lambda ^{ N }
\end{equation}
\noindent
The usual Poisson factorial does not appear
because the number of counts in a tick is 0 or 1.
Marginalizing $\lambda$ 
using the conjugate Poisson prior\index{conjugate prior}\index{Poisson}
(\cite{gelman}; Section 2.7, p. 49-50)
\begin{equation}
p(\lambda) = {\beta^{\alpha} \over \Gamma(\alpha)}
e^{-\beta\lambda} \lambda^{\alpha - 1} \ ,
\end{equation}
\noindent
the contribution to the Bayes factor for a cell -- 
often called the {\it marginal likelihood} -- is 
\begin{equation}
P( M | D ) =
\int_{0}^{\infty} p(\lambda) L( N | \lambda, V ) d\lambda = 
{\beta^{\alpha} \over \Gamma(\alpha)}
{\Gamma(N +\alpha) \over (V+\beta)^{N+\alpha} } \ .
\label{marg_like_1}
\end{equation}
\noindent
Based on a prior assigning a uniform distribution
to the probability of occurrence of a single event,
the marginal likelihood 
\begin{equation}
P( M | D ) = 
{ \Gamma(N+1) \Gamma(V-N+1) \over \Gamma(V+2) } \ ,
\label{marg_like_2}
\end{equation}
\noindent
was derived in \cite{scargle_v}.
A function of the same two sufficient statistics,
$N$ and $V$, 
it behaves similarly  to the
marginal likelihood in Eq. (\ref{marg_like_1}), but was found
less accurate in simulation studies
such as the one discussed below.

\section{Three Approaches to the Multiple Changepoint Problem}
\label{appr_3}

We now have to evaluate the above marginal likelihood
for each segment of a
piecewise homogeneous Poisson
model consisting of successive, 
independent\footnote{That is, 
due to a fundamental property of the Poisson distribution,
random variables corresponding to counts in
successive blocks 
(or different cells in general) 
are independent.
This fact should not be confused with
independence of the actual Poisson rates,
which in general does not hold.} 
blocks of data, and cobble the results together. 
The next section describes 
three ways to do this.
The first two seem effective in 1D,
but have problematical extensions to
higher dimensions.  
The third was inspired by its simplicity in $2D$ and $3D$.
All three methods are
demonstrated on the same synthetic $1D$ data.

\subsection{An Iterative Approach: Top Down}
\label{top_down}

\begin{figure}[htb]
\hspace{6.5in}
\vspace{7in}
\par
\hskip -1.2in
\special{dandc.epsc2}
\caption{Block representations 
based on the divide and conquer algorithm.
The piecewise homogeneous Poisson 
data were generated
from a modification of Donoho's
Blocks function.
The actual changepoints used to generate
the data are shown as vertical lines
at the bottom, and the changepoints
determined on the first three steps of 
the iteration are dashed lines.}
\end{figure}

As detailed in \cite{scargle_v},
the formulas above allow easy 
computation of the Bayes factor comparing 
(1) the two-block model in which the interval is segmented
into two parts at a changepoint\index{changepoint},
with 
(2) a single Poisson\index{Poisson} model for the whole interval.
The decision whether 
to keep an interval unsegmented
or to divide it into two subintervals
is based on comparison of the corresponding marginal likelihoods.  
Let $\Phi(N,V)$ 
stand for the marginal likelihood 
corresponding to a Poisson\index{Poisson}
model for a volume of size $V$ containing $N$ events, 
such as one of the two functions given above.
Then an interval should be broken in two if
\begin{equation}
\Phi( N_{i}, V_{i} ) \Phi( N_{i+1}, V_{i+1} ) - \Phi( N, V ) > 0 \ ,
\label{divide_criterion}
\end{equation}
\noindent
where the putative changepoint\index{changepoint}
divides the interval of size $V$
into two parts, of size $V_{i}$ and $V_{i+1} = V = V_{i}$,
containing $N_{i}$ and $N_{i+1} = N - N_{i}$ events,
respectively.
This criterion is easily computed as a 
function of the location of the changepoint\index{changepoint}.
The interval is then segmented at the point
that maximizes the expression in 
Eq. (\ref{divide_criterion}).
This {\it divide and conquer}
scheme is applied first to the
whole data interval,
and then iteratively to any subintervals
found at the previous step.
When this criterion favors 
segmentation of no further intervals, 
computation halts.

Figure 1 shows the results
for synthetic data generated by 
a piecewise constant Poisson\index{Poisson} process,
between eleven known changepoints\index{changepoint}.
I used a modified form of the
{\it Blocks} function popularized by David Donoho,
Iain Johnstone, and the 
{\bf WaveLab} project\cite{buckheit_donoho}
as a sample function with many discontinuities
that can be detected using wavelet\index{wavelets}
methods (see Fig. 1 in \cite{ogden_5}).
The original function has blocks
of negative amplitude, which will not do as
Poisson\index{Poisson} rates.  Hence I added a constant,
$3{1  \over 2}$, to the classic Blocks function, and then 
generated points obeying the Poisson\index{Poisson} distribution.

The three panels show three divide and conquer iterations.
It can be seen that the various changepoint
locations are found rather accurately,
allbeit with the single changepoint algorithm.
In the early steps the posterior has multiple 
peaks at the various changepoints, but only the
highest one is used at each step.
There is a tendency for changepoints 
determined early in the process to be
less accurate.
These errors are not corrected as the iteration proceeds,
but the algorithm could be modified to do so.

\subsection{Maximum Posterior for $N_{cp}$ {\it via} MCMC}
\label{middle}

\begin{figure}[htb]
\hspace{6.5in}
\vspace{7in}
\par
\hskip -1.2in
\special{mcmc.epsc2}
\caption{Block representations 
based on the MCMC algorithm.
The piecewise homogeneous Poisson 
data were generated
from a modification of Donoho's
Blocks function.
The actual changepoints used to generate
the data are shown as vertical dashed lines.
Spurious, narrow blocks have been emphasized
({\it e.g.} the one at about $t=0.76$.}
\end{figure}

It is clear that the above method is not rigorous,
in that it does not solve for all of the
changepoints\index{changepoint} simultaneously.
It is relatively straightforward,
however, to remedy this with
a more rigorous, but also more computationally
intensive scheme.
If there are multiple changepoints, 
say $N_{cp}$ in number,
the full posterior is just the product of 
the posteriors of all the subintervals.
The marginalization of all the locations
of the changepoints 
requires evaluating the $N_{cp}$-dimensional
integral\footnote{In practice, this is
a finite sum, {\it e.g.} over a bin index or an event index.}
of the posterior over all 
values of the changepoint\index{changepoint} locations.

Results obtained in this way,
using a simple Markov Chain Monte Carlo (MCMC) method,
are surprisingly good.
Figure 2 shows the block representation
of the same data as in Figure 1, 
obtained from a simple MCMC
evaluation of the marginal likelihood
as a function of the number of changepoints.
The maximum posterior was at 16 changepoints
(17 blocks), compared to the correct
value of 11.  The ``extra'' blocks
are narrow, and while they do not
look pretty, they will not much affect
derived quantities (such as pulse widths,
rise times, {\it etc}).

The good performance
with only a few iterations
may be due to a combination of several factors:
\begin{itemize}
\item[$\bullet$] the well behaved nature of the data 
   \begin{itemize}
   \item[$\circ$] low dynamic range of the signal
   \item[$\circ$] well behaved backgrounds
   \item[$\circ$] exact Poisson\index{Poisson} statistics
     for the observational uncertainties
   \end{itemize}
\item[$\bullet$] a simple, exact Likelihood; only one parameter
\item[$\bullet$] the fact that the posterior does not
have to be computed accurately to
distinguish one value of $N_{cp}$ from another
\item[$\bullet$] the useful characteristics of the final model
   are not that sensitive to $N_{cp}$
\end{itemize}
\noindent
Nevertheless, when the number of changepoints\index{changepoint}
and the number of data points
are large, the computation is quite long.
Since there is little global communication,
in the sense that the location of a change
point in one part of the observation 
interval affects that of one elsewhere,
breaking the data up into smaller intervals
is an effective way to speed up the overall
computation.

\subsection{Cell Coalescence (Bottom Up)}
\label{bottom_up}

\begin{figure}[htb]
\hspace{6.5in}
\vspace{7in}
\par
\hskip -1.2in
\special{voronoi.epsc2}
\caption{Block representations
using the Cell Coalescence algorithm.
The top panel shows the overly fine
segmentation in the early stages of the iteration.
In successive panels the process is converging
toward a coarser representation.
The bottom panel shows the first stage
at which the Bayes factor\index{Bayes factors}
contraindicates merging of all of the remaining blocks.
The actual changepoints used to generate
the data are shown as vertical dotted lines.}
\end{figure}

Based on preliminary tests, 
a new algorithm may be a significant
improvement over either of the above approaches.
It begins with a fine-grained segmentation,
namely the Voronoi tessellation of the data points,
and merges these many cells to form fewer, larger ones.
The outline of the algorithm is simple:
\vskip 0.25in

\noindent
{\fboxsep=0.25in \framebox{\parbox[t]{4.75in}
{
\begin{center}
Bayesian Cell Coalescence
\end{center}
\begin{itemize}
\item[$(0)$] Initial Segmentation: Voronoi Tessellation of the Events
\item[$(1)$] Compute Bayes Factor\index{Bayes factors}
for Merging Each Pair of Adjacent Cells
\item[$(2)$] Identify Largest Bayes Factor (at $i$)
\item[$(3)$] If Largest Bayes Factor is $<1$, HALT!
\item[$(4)$] Otherwise Merge Pair of Cells with Largest Bayes Factor:
   \begin{itemize}
   \item[$\bullet$] $N_{i} \leftarrow N_{i} + N_{i+1}$
   \item[$\bullet$] $V_{i} \leftarrow V_{i} + V_{i+1}$
   \item[$\bullet$] Delete Cell $i+1$
   \end{itemize}
\item[$(5)$] Go to $1$
\end{itemize}
}}
\vskip 0.25in
\noindent
While there is no rigorous justification 
for this procedure,
one has the sense that it should
come reasonably close to obtaining 
the optimal solution.
At each step in the iteration 
the local event rate in a cell is 
${N \over V}$, 
the number of events in the cell divided by its volume.
The rate estimates for the initial cells,
${1 \over V}$, are reasonable estimates 
of the fine-grain, local event rate
if these cells 
are assigned roughly their surrounding volume, 
extending approximately halfway to neighboring points.  
An obvious choice for the initial partition 
is thus the 
{\it Voronoi tessellation}\index{Voronoi tessellation} 
\cite{stoyan,klein} of the data points.
The Voronoi cell for a data point consists of 
all the space closer to that point than to
any other data point.

The Voronoi tessellation of a
set of points on an interval (1D)
is trivial: it is simply the 
set of intervals spanned by 
pairs of midpoints between successive data points.
Many algorithms exist for computing 
Voronoi tessellations in
higher dimensions\cite{nolt,edel,preparata,deberg},
allowing one to address
problems such
as cluster detection in parameter
spaces, and identification of structures in images.
Since the marginal posteriors discussed above
are valid for piecewise constant Poisson 
data in any dimension, 
the methods demonstrated here for $1D$
should apply in general.  

The decision  
whether to merge two cells or to halt,
Step (3),
is based on comparison of the Bayes factors.  
Using the same notation
as above, in Eq. (\ref{divide_criterion}),
cells $i$ and $i+1$ are merged if
\begin{equation}
\Phi( N_{i} + N_{i+1}, V_{i} + V_{i+1} ) - \Phi( N_{i}, V_{i} ) \Phi( N_{i+1}, V_{i+1} ) > 0
\end{equation}
\noindent
and kept separate otherwise. 
When this criterion favors the
merging of no further cells, 
computation halts.

Figure 3 demonstrates
the application of this algorithm
to the same synthetic data
as in the previous two examples.
The top panel depicts the
state part way into the
iterations, but is not far from the
initial Voronoi\index{Voronoi tessellation} tessellation,
one Voronoi cell for each data point.
Successive panels show the
evolution of the representation
as the cells merge into fewer, larger
ones.
In the last panel, 
the halting criterion mentioned above
has just terminated the iteration.
It can be seen that most of the
changepoints\index{changepoint} 
are accurately detected.  
Several though are missed.

%========================================
\section{Related Work} 
\label{related_work}
%----------------------------------------

It is well known that
Bayesian methods are well-suited to 
finding changepoints\index{changepoint}
({\it e.g.} \cite{ruan,raftery_akman};
see also Appendix C of \cite{gregory_loredo}).
In \cite{west_ogden} methods 
similar to those described here 
are used 
to find changepoints\index{changepoint} in binned data,
to an accuracy better than the bin size.
A number of recent publication are 
relevant to this problem
\cite{stark,gustafsson_1,gustafsson_2,west_ogden,carlin,raftery_akman,stephens_1,stephens_2,pettitt,chib}
and \cite{ogden_0,ogden_1,ogden_2,ogden_3,ogden_4,ogden_5,ogden_6,ogden_7}.
More recently,
Raftery and colleagues have
developed similar methods,
mainly for $2D$ problems.
In \cite{byers_raftery} 
the Voronoi sites are fiducial markers for the cells,
not the tessellation defined by the individual data points.
Other approaches are described in 
\cite{fraley_raftery,byers_raftery_2,dasgupta_raftery,walsh_raftery,raftery}.
% An outline and notes for a course in Bayesian Inference,
%in the Department of Mathematics and Statistics,
%ylde College, Lancaster University, Lancaster, England,
%given by Stuart Coless\cite{coless}
%discusses 
%\verb+http://www.maths.lancs.ac.uk/~coless/btch/btch.html+;
%\verb+http://www.maths.lancs.ac.uk/~coless/btch/node67.html+
%351 - Statistical Inference.
Bayesian methods are also useful for segmentation
of autoregressive models, with applications to speech
processing \cite{cmejla}.

After this article was completed,
I became aware that NASA's 
Chandra X-ray Observatory, is
using an unbinned source detection technique 
much like cell coalescence, 
and starting from
the same Voronoi tessellation\cite{ebeling}.
Source cells are merged with a
{\it percolation} process
based on a sample estimate of the
density distribution.

%===============================

\section{Acknowledgements}

I am greatful to 
Alanna Connors,
Seth Digel,
Tom Loredo, 
and
Jay Norris
for helpful comments and encouragement.
This work is partly funded by the
NASA Applied Information Systems Research Program.

\end{document}